# Design and Implementation of a Complete Wearable Smart Insole Solution to Measure Plantar Pressure and Temperature

Sakib Mahmud[1], Amith Khandakar[1,2], Muhammad E. H. Chowdhury[1*], Mamun Bin Ibne Reaz[2*], Serkan Kiranyaz[1], Zaid Bin Mahbub[3], Sawal Hamid Md Ali[2], Ahmad Ashrif A Bakar[2], Mohammed Alhatou[4], Mohammed AbdulMoniem[1]

[1]Department of Electrical Engineering, Qatar University, Doha 2713, Qatar

[2]Department of Electrical, Electronics and Systems Engineering, Universiti Kebangsaan Malaysia, Bangi, Selangor 43600, Malaysia

[3] Department of Physics and Mathematics, North South University, Dhaka, Bangladesh

[4] Neuromuscular Division, Hamad General Hospital and Department of Neurology; Alkhor Hospital, Doha, 3050, Qatar

∗ Correspondence: Muhammad E. H. Chowdhury (mchowdhury@qu.edu.qa); Mamun Bin Ibne Reaz (mamun@ukm.edu.my)

**Abstract:** A complete smart insole solution that continuously monitor the foot plantar pressure and temperature can detect foot complications early and that too from the convenience of the user's home. Widespread health complications such as Diabetic Mellitus need continuous foot complication monitoring to avoid severe complications. With that motivation, this paper provides a detailed design of a wearable insole using popular off the shelf sensors to monitor foot plantar pressure and temperature. The design provides details of which temperature and pressure to be used, circuit configuration for characterizing the sensors, the considerations for developing a compact Printed Circuit board design using appropriate microcontroller and communication protocol. The system also provides details of how the foot pressure and temperature data from the subjects using the sensors can be transferred wirelessly using a low-power consuming communication protocol to a central device where the data will be recorded. The investigation can help in developing low-cost, feasible and portable foot monitoring system for patients by facilitating real-time, home monitoring of foot condition using Gait Cycle or Foot Pressure patterns and temperature heterogeneity between two feet. . The proposed system will work in real-time.

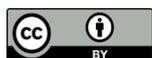

## 1. Introduction

The advancement of electronics and sensors has helped in the development of a more portable and reliable diagnostic wearable for the early detection of anomalies [1]. The emphasis on e-Health, especially during the pandemic period, has driven the need for self-diagnosis solutions (using sensor systems with machine learning solutions) and help the medical staff in the early diagnosis of anomalies without compromising their daily routines [2-4]. Amongst the various health complications needing special attention, Diabetes Mellitus (DM) is a chronic medical condition resulting from a high amount of sugar in the blood, which often leads to other severe health complications like heart-related diseases, kidney failure, blindness, and lower limb amputation [5]. Diabetes is known to cause neuropathy, especially in the feet, and can result in incurable infections. This can lead to foot ulcers which are localized lesions in the skin, or underlying tissues. This is caused due to difficulty or delay due in blood circulation to the infectious area [6].

A diabetic patient under the 'high-risk' category requires regular check-ups, hygienic personal care, and continuous expensive medication to avoid unwanted consequences. Diabetic foot ulcers lead to increased healthcare costs, decreased quality of life, infections, amputations, and death. Early detection and better classification tools for DFU symptoms can enable correct diagnosis, effective treatment, and timely intervention to prevent foot ulceration, amputation, and death.

Self-diagnosis at home, i.e., self-care, which means monitoring without medical assistance, could be useful in preventing severe after-effects in the case of DFU. However, the easiest monitoring technique, visual inspection, has its limitations such as people with obesity or visual impairment not being able to see their sites of ulcer easily. Another effective technique could be monitoring the feet' temperatures regularly, which can be useful as an early warning system. This achievable technique can provide patients with feedback and alert them to adjust their activity to prevent ulcer growth. International Working Group on Diabetic Foot clinical practice provides guidelines on such monitoring [7]. According to recent studies, a temperature monitoring system at home has been able to spot 97% of diabetic foot ulcers well in advance [8-10]. It has also been confirmed that patients going through continuous monitoring of the feet temperature had a low risk of foot complications [8, 11]. Similar to temperature difference, the plantar pressure also provides information regarding diabetic foot complications. Nahas et al. in [12] have confirmed the strong correlation of plantar temperature and pressure distribution in a study involving 25 healthy patients. Later, Deschamps et al. in [13], with the help of statistical analysis and clustering of plantar pressure maps using k-mean clustering, confirmed the distinguishability between Healthy and Diabetic patients. Thus, improved sensors to provide these temperature and pressure maps can help in monitoring the progression of diabetic foot complications.

Several tools to measure plantar temperatures are available with the limitations of allowing temperature measurement only once a day, or they are designed for use only under clinical supervision [14]. One of the solutions used an infrared, handheld thermometer for measuring the temperature at six locations of both the feet each morning and comparing them [15, 16]. The threshold temperature difference of 4°F (2.22°C) or higher was used as a reference for the early sign of DFU. However, this tool can also result in false alarms as it can be subjective to manually measure the temperature at different locations of the foot, especially when the feet have different sizes and shapes [8]. Another solution used a "smart mat" for measuring the foot temperature daily [16]. The nature of the temperature variations on both of the feet could be used to find the locations with higher temperatures and thus identify the formation of potential ulcers at an initial stage. There are a few other innovative wearables such as "smart socks", "smart Insole" with embedded sensors for measuring temperature [8, 11, 17]. There are socks made entirely of optical fiber [18], which showed great promise but has a huge drawback due to the fragility of the optical fiber while wearing the sock [16]. Another report described electronic socks which measure the temperature of feet every 10 minutes and report to a mobile application [8]. Bluetooth-enabled socks with embedded sensors were designed for continuous, regular temperature monitoring in a home environment [19]. While some of the solutions are not suitable for real-time monitoring, others are washable socks but made up of fragile fibers, and some only track either temperature or pressure. Plantar pressure is also important to avoid false temperature asymmetry alarms. Moreover, the socks-based design has the issue of reusability for a longer duration. There are many sensors and electronic devices available for acquiring plantar pressure using piezoelectric and/or piezoresistive sensors and capacitive sensors [20, 21].

The solutions are either expensive or does not provide a complete solution for monitoring Foot Plantar Temperature and Pressure. To address all the issues with available options of wearable foot complication detection, in this paper, we propose and compare different sensor-based wearable smart footwear for continuous plantar temperature and pressure measurement systems, which will measure the daily asymmetric temperature and pressure variations due to the gait dynamics for early, reliable, and robust detection of diabetic foot complication. The captured temperature and pressure data generated throughout several full gait cycles can be sent to the PC module, logged into a database for future analysis, and can also be used to synchronously update the gait dynamics. In this way, the user can get an early warning and can contact health care professionals to take preventive measures before the disease reaches chronic conditions.

The paper is divided into 4 sections which are Section I reflects the motivation behind the study and some of the recent works done in this area while Section II discussed the research methodology with the details of the sensors and how they were characterized, details of the microcontrollers and communication protocol investigated in the study, Section III provides the results from the different investigations done in the paper , followed by Section IV providing discussion of the results from the investigation and presenting the final solution. Finally the conclusion is provided in Section V on how the proposed solution can be used for remote plantar health monitoring.

**2. Methodology and Experimental Details**

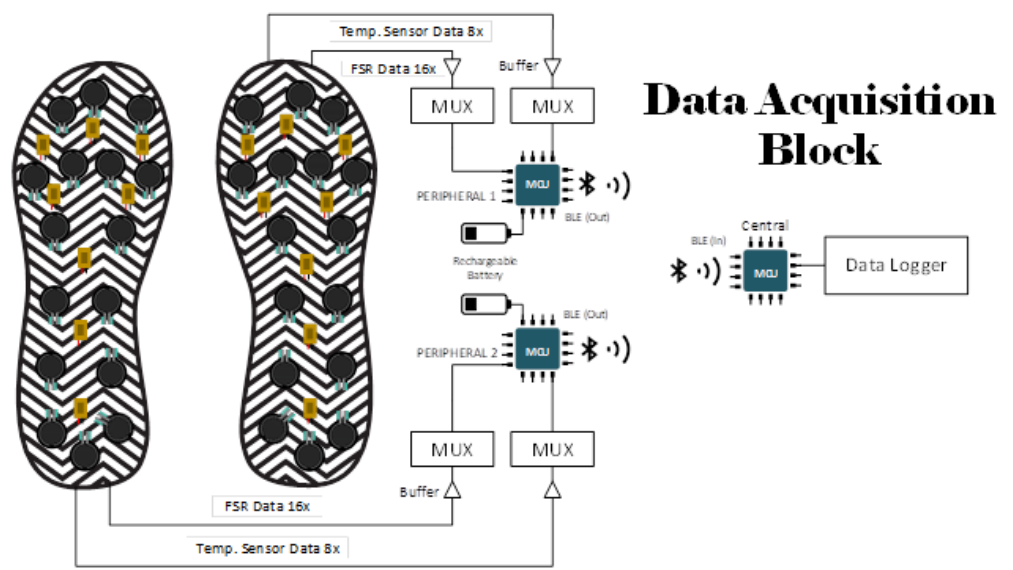

Figure 1: Major Blocks of the Study

This section will talk in details about the various popular off the shelf and readily available sensors for plantar pressure and temperature monitoring, microcontrollers and communication protocol. The block diagram of the whole data acquisition system is shown in figure 1, which consists of the footsole, multiplexer to combine the plantar pressure and temperature signals, microcontroller with power source and communication sub system.

## 2.1 Pressure Measuring Sensors

Three popular off the shelf sensors were investigated in this paper for smart foot sole purpose i.e. Velostat Sheet, Force Sensitive Resistors (FSRs), Piezoelectric Sensors. Figure 2 is showing the placement of these three sensors on the Insole, the placing of the sensors are based on the investigation done by the authors in their previous study and from literature [2, 20, 22, 23].

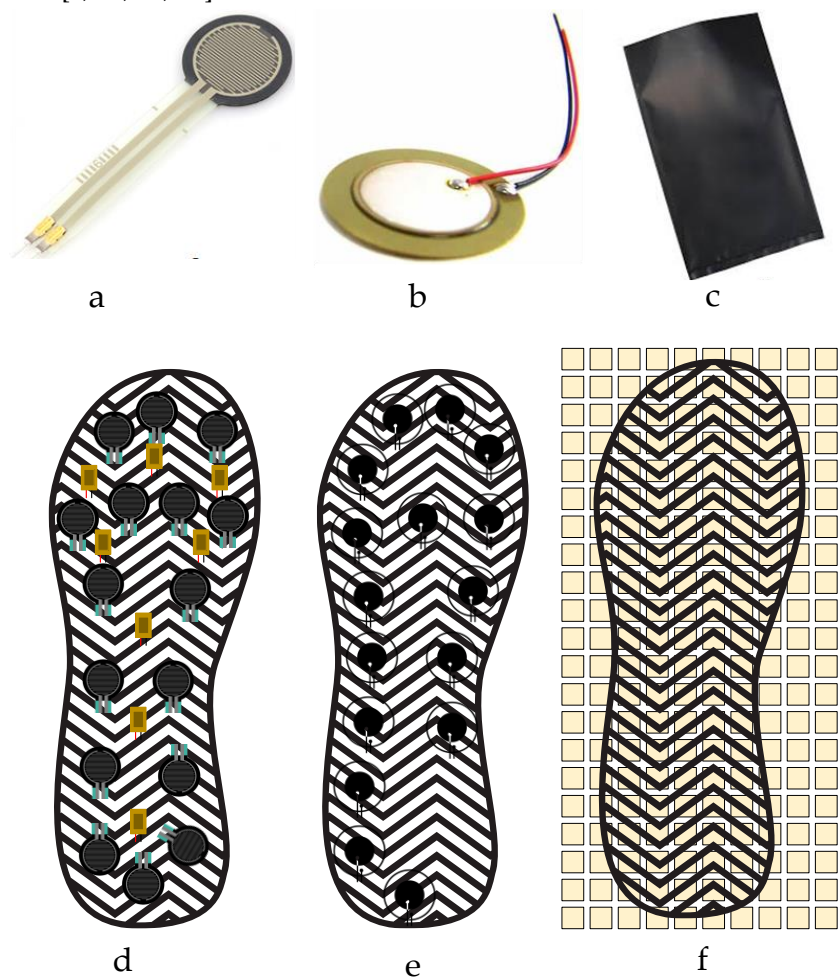

**Figure 2:** Three different electronic pressure sensors and their placements: **(a)** Velostat; **(b)** Piezo-electric; and **(c)** FSR **(d)** FSRs (e) Piezoelectric Sensors (f) the Velostat Sheet (with switching matrix) on an Insole

*Velostat Characterization*

The velostat sheet is used to measure the applied pressure on the foot according to the variation of resistance value, as the pressure increase the resistance must decrease. A load characterization test was implemented to check the response of the sheet to different weights and how the resistance varies so that it can be used to analyze vertical Ground Reaction Forces (vGRFs), which represents the magnitude and pattern of mechanical loading in the vertical direction at the foot, and can be commonly used in the diagnosis of atypical gait. The test was about placing the velostat sheet between two conductive sheets to start loading the velostat by light loads and measuring the resistance by multimeter, then the loads must be removed to see the response and the readings of the sheet without the applied load. The loads cannot be placed directly on the conductive sheets that is why two plastic plates were added as shown in Figure 3.

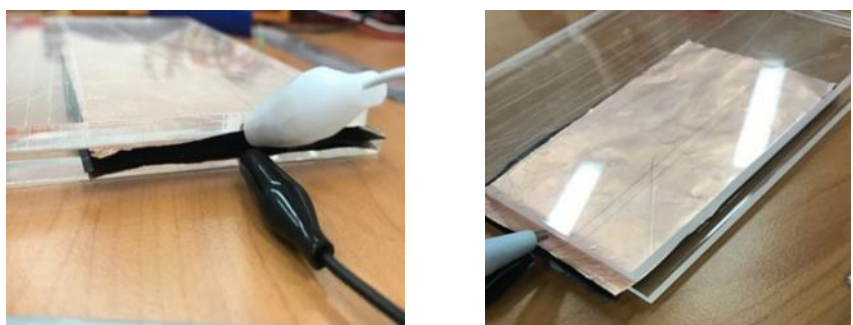

Figure 3: Characterization setup of the Velostat Sheet

*FSR Characterization*

Before starting the characterization of FSR, a solid plane must be provided to place the FSR on; to avoid deformation of the FSR. Moreover, the weights have to be applied only on the active area of the FSR and not touching any other surface to assure that the whole weight is applied on the sensor only; consequently, a 12.7mm acrylic cylinder along with a plastic rectangle were placed above the FSR active area to support the weights, as shown in figure 4.

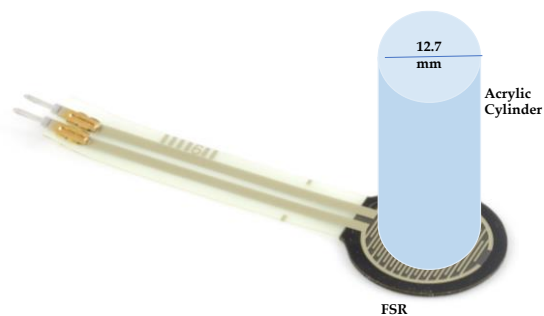

Figure 4: Characterization setup of the FSR system

*Piezoelectric Characterization*

This sensor is measuring dynamic pressure only and thus had to be characterized using other method than applying weight gradually to measure if the voltage is constant based on applying constant pressure. Mainly, piezoelectric sensor measures the existence of dynamic force, it does not measure the force quantity [24]. A vibrational continuous monotonous force test was done to characterise the sensor. As shown in Figure 5, the piezoelectric sensor was connected to a parallel 1MΩ resistor, grounded from one pin and the other pin is connected to analog input of the MCU to measure dynamic voltage at 1kHz sampling frequency. The code was written using the Arduino software where it starts by defining the input variable, then the software starts reading from the MCU the peak input voltage only and plot it and a delay has been used in order to avoid any over loading in the serial port.

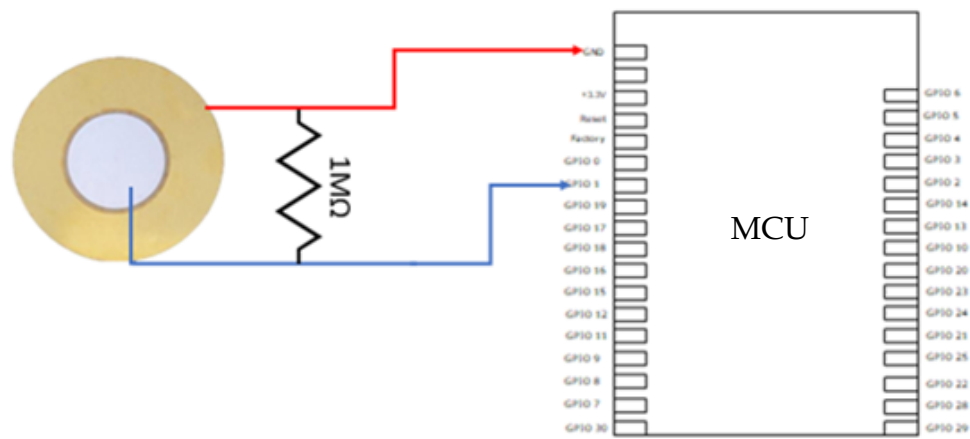

Figure 5: Characterization setup of the Piezoelectric Sensor

## 2.2 Temperature Measuring Sensors

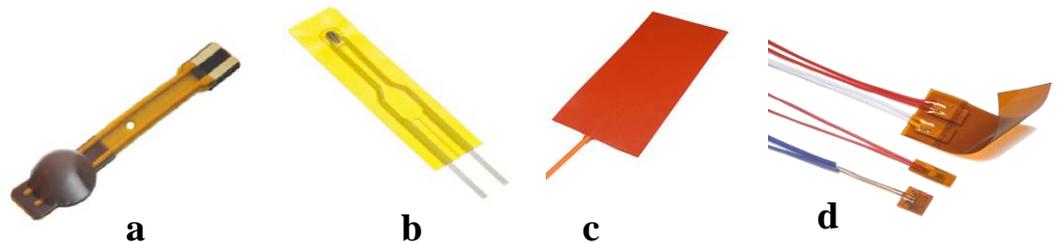

**Figure 6:** Three different electronic temperature sensors: **(a)** 10k NTC Thermistor module on Flex Cable; **(b)** 10k NTC Thermistor; **(c)** Flexible Thermostat heat pad, and **(d)** Thermal-Ribbon™ Flexible RTD [25]; which are tested for the insole application

To study the temperature of the foots, different temperature sensors have been analysed in this research. Three types of commonly used temperature sensors have been considered for this purpose, they are: Thermostats, Resistance Temperature Detectors (RTDs), Thermistors.

Among them, thermistors were chosen due to some important reasons:

• The most important reason was that the Thermistors could be hooked up easily using the same voltage divider similar to the other pressure sensors and will make the circuit design for PCB lot easier and helped in keeping its size minimum. On the contrary, RTDs require a Wheatstone Bridge for each of the sensors. Placing eight bridges on the PCB would drastically increase its size, which is also a critical constraint. The thermostats are not suitable for this application due to their nature of work.

• The foot temperature of a person can vary from 15°C (when the feet are cold during cold season) to about 37.5°C during the summer (the study was done on normal people when they were awake) [26]. Since the aim is to find the temperature of the feet, the suitable working range of the temperature sensor was chosen as 20°C to 50°C. Selecting this range was crucial for the calibration of the sensor. Now, Thermistors work very well within this range, and their response is close to linear.

- Thermistors have very low response time, i.e., responses very fast, faster than RTDs and Thermocouples, which is critical for real time application.

Selecting Thermistors as the temperature sensor was not enough since thermistors (and any other sensor) comes into various packaging differed by size, shape, colour, and advanced options like packaging material, water/thermal proofing, etc. For the insole design, the required properties of the temperature sensor module should be as follows:

- Flexibility: Firstly, the sensor will be embedded into the insole, which will be worn by a person under test. So, the sensor module must be flexible to ensure that the insole is not creating any impedance on the subject while walking (which might affect the gait cycle and provide biased data).
- Small Size: The sensing module should be small since temperature readings from various points under the feet need to be taken. Using sensors with the big surface area will not only block the pressure sensor but also will fail to provide the temperature data of a small section.
- Durability: The sensor module should be durable enough along with its flexibility since the insole will be worn inside a shoe. It needs to withstand shear, lateral, and vertical pressure applied to it while walking. So, the sensor module should have a certain degree of durability, not too fragile. It is also indirectly related to flexibility (to withstand shear pressures and bends).
- Temperature Range: Temperature range should be able to contain the peak temperature readings from feet under any condition.
- Waterproofing: The sweating occurring (for some people it is more acute) in the leg after wearing the shoe for more than a few minutes is tackled by using waterproof temperature sensor modules.

The temperature response of the sensor module need not be very fast or dynamic like the pressure sensor since the temperature record need not be instantaneous. But having a fast response and a high sampling rate, if it does not overwhelm the hardware is always a plus point.

### 2.3 Communication Protocol

Under IEEE 802.11, which is, in turn, a subset of the IEEE 802 Local Area Network (LAN) standard, there are several protocols that have been used in various applications for the last two decades. Any of these 2.4 Ghz frequency-based protocols like ZigBee, Wi-Fi, or Bluetooth can be used for wireless data transmission. The selection criteria followed while choosing the suitable protocol for data communication are discussed below in brief:

- **Bandwidth:** Bluetooth (both traditional and LE) uses the Industrial, Scientific, and Medical (ISM) band of 2.4 GHz, which makes it suitable for medical usages [27]. Bluetooth also has good immunity to noise, making it less susceptible to noises made by medical devices, which is less in case of Wi-Fi. Bluetooth uses a modulation technique called Adaptive Frequency Hopping (APH) in which it switches between several frequency channels along with its bandwidth and avoids busy channels. This helps Bluetooth to minimize interferences created by other devices running on Bluetooth or any other protocol under the same band [28].

- **Power Consumption:** The peripheral devices needs to be portable for this application since they are attached to the subject's leg, connected to the smart insole. In this case, the battery consumption rate is a critical issue since if the battery depletes fast, either it will have to be changed frequently or be recharged if they are rechargeable. The power consumption rate of various RF based Wireless Personal Area Network (WPAN) protocols were compared in a background study to select the best protocols suitable for this purpose.

Table 1: Comparison Chart of Various Common RF Based WPAN Protocols

| Properties | Bluetooth LE (v4.2) | Bluetooth | Wi-Fi | ZigBee |
|---|---|---|---|---|
| IEEE Standard | 802.15.1 | 802.15.1 | 802.11a/b/g/n/ac | 802.15.4 |
| Bandwidth | 2.4 GHz | 2.4 GHz | 2.4 or 5 GHz | 868 MHz - 2.4 GHz |
| Maximum Data Transfer Rate | 1 Mbps | 1-3 Mbps | 450 Mbps (802.11n) | 250 kbps (2.4 GHz) |
| Range | 10–40 m | 10–100 m | 150 m (2.4 GHz) | 10-100 m |
| Average Power Consumption | Low | Medium | High | Low |
| Power Consumption over Data Transfer | Medium | Medium | Low | High |
| Latency | 2.5 ms | 100 ms | 1.5 ms | 20 ms |

From Table 1, it can be witnessed that Bluetooth LE is a Low-Power communication protocol, among others. But the exact power consumption depends on many factors, e.g., data transfer rate (which depends on power level), chip latency antenna (receive and transmit) sensitivity and gain, etc. For example, the nominal range of BLE was mentioned as 10 – 40m, as mentioned in Table 1. But this is related to the nominal data transfer rate (1 Mbps) used by most chip manufacturers. If someone wants to increase the range, maintaining the same data transfer rate in the same environment, more power will be required. On the other hand, increasing the range and maintain low power consumption will feed on the data transfer rate (which is exactly being done in the physical layer of Bluetooth 5 protocol) [29]. Transfer power allowed by the Bluetooth Protocol ranges from -20 dBm to +20 dBm, which allows a range from less than a meter to more than a kilometer [30] But most chip manufacturers do not (or cannot) go up to that level due to many constraints they face during the design phase. For example, Nordic's nRF52832 can output up to +4dBm, while the nRF52840 can output +8dBm [28].

- **Range:** The working range of the communication protocol is also an important parameter to judge while selecting. The range of a particular module depends not only on the communication protocol chosen but also on parameters like the sensitivity of the receiver module, gain of the transmitter and the receiver antenna, transmit power and path loss (which depends on the surrounding).

**2.4 Microcontroller**

The selection of the Microcontroller Unit (MCU) was a tedious task since there are many MCU modules available in the market which can work equally efficiently in making Bluetooth Low Energy communications. Some of the candidates were esp32 from Espressif [31], Adafruit Huzzah/Feather [32], MCUs from nRF semiconductor [33], Arduino Nano BLE Sense [34], etc.

Among all these options, esp32 was chosen for the following reasons,

- Esp32, unlike most other MCU (except original Arduino modules), has one of the most active developer bases on the web. There are so many people who are creating BLE and Wi-Fi-based projects with esp32. Due to the active, current, and helpful online community esp32 has, it was chosen as the MCU.
- MCUs who offer BLE facilities run on various Bluetooth versions, the latest one being Bluetooth 5. .Esp32 was running on Bluetooth 4.2 which has a significant benefit over the 4.0 version due to its ability to deal with a much bigger packet size. Notable that Espressif has already been planning to make Bluetooth 5 available in esp32 chips.
- The BLE library developed for esp32 is the best among the MCUs available in the market due to its versatility, ease of understanding, expandability, so on and so forth. Developing such a library requires great knowledge in the domain of software development, and this library developed by Neil Koblan is really a piece of art [35].
- Esp32 has many types of developing modules, which are, in turn, based on various chips from esp32. To reduce the PCB size as much as possible, the soliution was to look for the smallest esp32 kit available in the market. The comparison of the sizes of different modules of esp32 can be seen in figure 7.

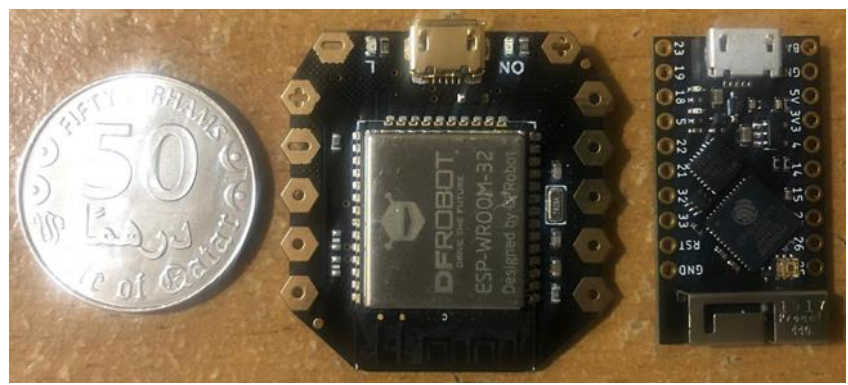

Figure 7. Beetle esp32 [36](left) and TinyPico [37] (right) compared to a 50 Dirham

The TinyPico was selected as the solution due to the amazing features like,

- TinyPICO has almost all the features of any other esp32 module (e.g., ADC, DAC, GND, RST, 3.3V PWR, 5V PWR, SCL-SDA, MOSI-MISO, so on and so forth) inside that tiny body of it.
- TinyPico manages the voltage output level and harmonic contents very well using the embedded Low-Dropout Regulator or LDO from the Texas Instruments.
- This LDO and some other intelligent circuitry by the developer and maker "UnexpectedMaker" allows the TinyPICO to be powered from an external source directly (in this case, it is a 3.7V Li-Po battery), in few ways. Other esp32 development boards also have this feature. But the most amazing thing about TinyPico is its "VBAT" pins. The VBAT (meaning Voltage Battery) pin can be used to charge a (rechargeable) battery. The Li-Po battery with the protection circuitry embedded was the perfect choice for this scheme.

- Compared to Beetle esp32, TinyPICO has three more significant advantages, which made the system much more efficient. Firstly, TinyPico comes with a 3D antenna (Beetle esp32 has a 2D PCB Printed antenna instead). 3D antenna provides a much better spectral efficiency, which makes the device work much better inside a confined place, especially if it is inside a metal container. Our system will be placed inside a box. Even if that will be made of non-metal plastic-type 3D printing materials, having an MCU with a 3D antenna will always be a bonus.
- Secondly, TinyPICO has more GPIO pins than Beetle esp32. Beete esp32 has four analog and four Digital GPIO pins. While TinyPICO has much more, as shown in Figure 7. To control the two mux used for the pressure and Temperature sensors, the MCU requires at least nine working GPIO pins. If Beetle esp32 was used, it would not be easy to implement temperature sensors to the insole.
- Thirdly, TinyPico has a regular MCU shape and breadboard-friendly, i.e., it can be used as a testing kit in a breadboard. The unusual shape of the Beetle esp32 (Figure 7) made not only the PCB design harder, but soldering, it was also a tedious task to do.

**2.5 Power Supply Unit**

The data acquisition device is planned to be portable and rechargeable. It will also be used for human usages, so it must have some sort of protection circuitry embedded. Therefore, the Power Supply Unit should roughly possess the following qualities:

- **Light:** It should not be heavy enough to impede the Gait Cycle of the subject. Since the PCB is already light and small, using a light power supply unit, which could be a battery, will save the day.
- **Small Size:** The size of the battery should not exceed the size of the PCB so much. Otherwise, the surface efficiency achieved by reducing the PCB size will go in vain. Lightness and small size will make the hardware much more portable, and suitable for the subjects.
- **Ample Charge Storage:** The device is planned to be portable. If the charge of the battery perishes very fast, the overall aim of portability will not be justified. So, a power supply unit with a small size but dense enough to store a good amount of energy should be used.
- **Safe:** The device will be used on a human body, so the safety factor must be assessed. A protection circuitry must be present along with the PCB to protect the device from overcurrent, overvoltage, so on and so forth.
- **Compatible and Easily Detachable:** The battery should be compatible with the hardware specifications and easily detachable for charging.

After assessing these basic requirements for the power supply units, the Lithium Polymer (Li-Po) batteries with embedded protection circuitry, has been chosen for the project. This battery has other additional benefits such as [38] the ability to supply high current (can deal with current peaks from BLE Tx/Rx), low self-discharge or charge leaking, possesses a high charge efficiency which allows the battery to be recharge and discharge for a long period of time without being dull (this is mainly due to the chemical efficiency of Li-Po). All these qualities increase the longevity of Li-Po and require low maintenance for the battery.

**2.6 Data Logging Unit**

The Real-Time Data Logger was developed so that the data received by the MCU acting as the Central BLE (i.e., the Data Acquisition System) can be saved permanently in a certain location (e.g., Folder) of the PC (i.e., Local Database) so that it can be used later on, refer Figure 8. The Data Logger was develop based on python since python being the easiest language to learn due to its high-level API

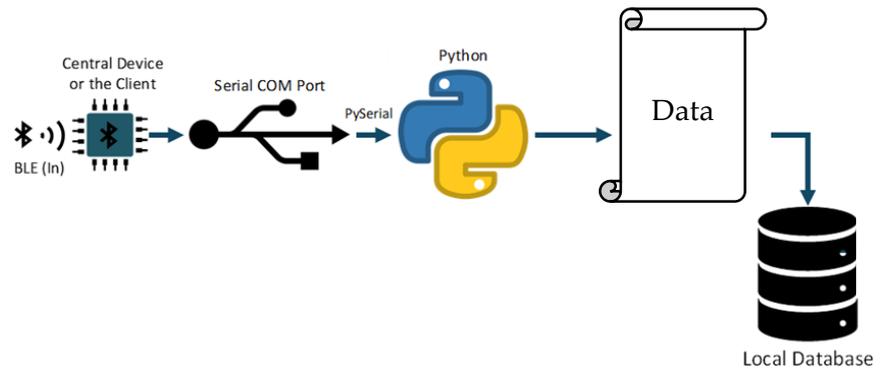

Figure 8: Block Diagram: Text-Based Data Logger

**2.7 Designing a 3D Box for Containing the Hardware**

A 3D Box was designed in Fusion360 to keep the PCB along with all its components and circuitry inside. It should be remembered that the device will be worn by a lot of different subjects from different backgrounds, and their idea about electrical circuitry might be multifarious, which might give rise to skepticism about our approach. So, the 3D box was designed, keeping in mind subtle issues like "Patient Perspective about the Device" and overall Aesthetics of the device. As can be seen from Figure 9, the box has been designed into two parts, which will attach to each other. The front side of the box has spaces for entering the ribbon cable (Figure 9(a)). There are spaces on the side of the box to make USB connections to the TinyPico (TinyPico uses micro USB-B to connect). The box was designed to almost fit the PCB perfectly, but there are some spaces for heat escape. A heat sink or a fan might be implemented in the future if needed. The box will get attached to the subject's leg(s) by passing a flexible VELCRO™ [39] through the bottom extension of the box, as shown in Figure 10. The flexible VELCRO™ has been tested and found to be very effective in keeping the box in place (i.e., not wobbly). Slight axial movements are fine for the setup dealing with pressure and temperature sensors.

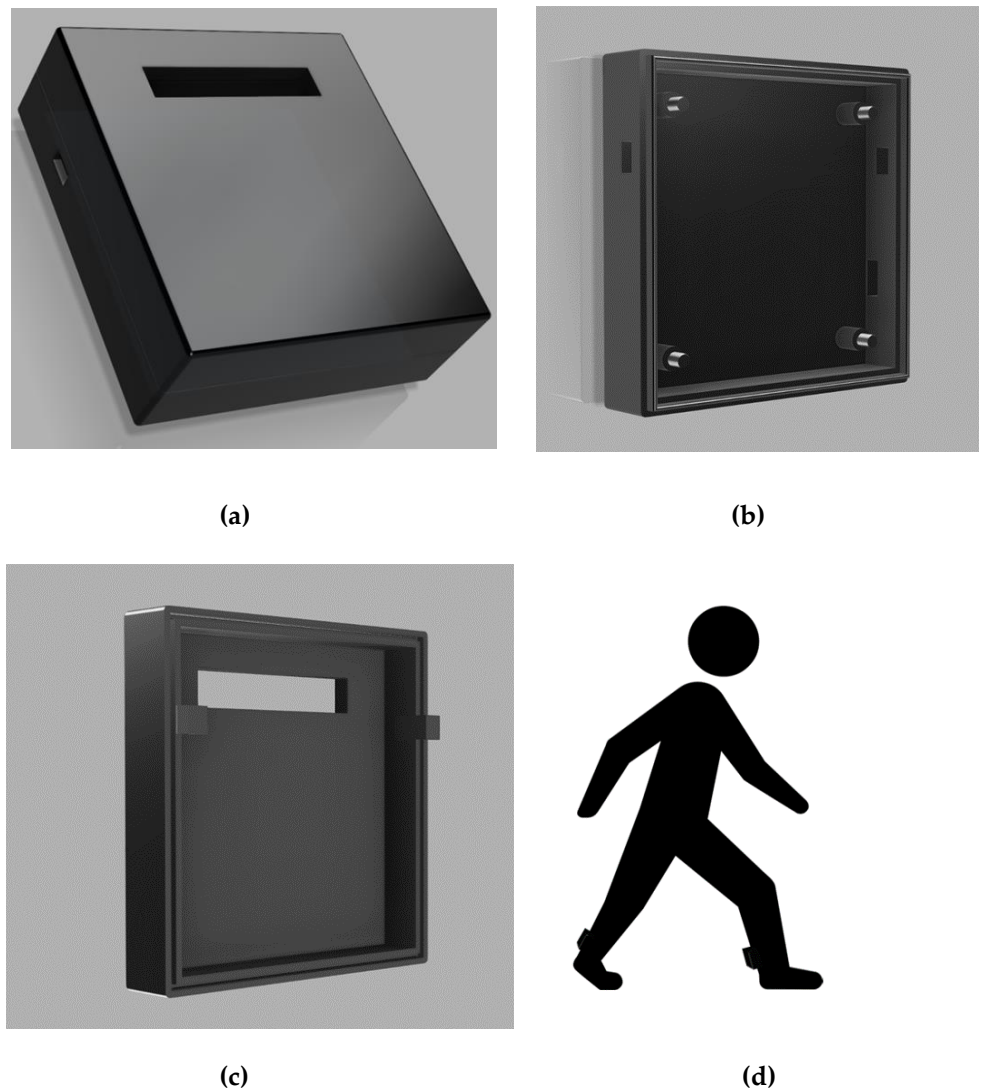

Figure 9: Rendered View of the Box (a) Full Box Closed; (b) Upper side of the Box Upside Down (From Inside); (c) Lower Side of the Box from Inside; (d) Model of a Person Walking with the Box and Medical Shoe for Visualization.

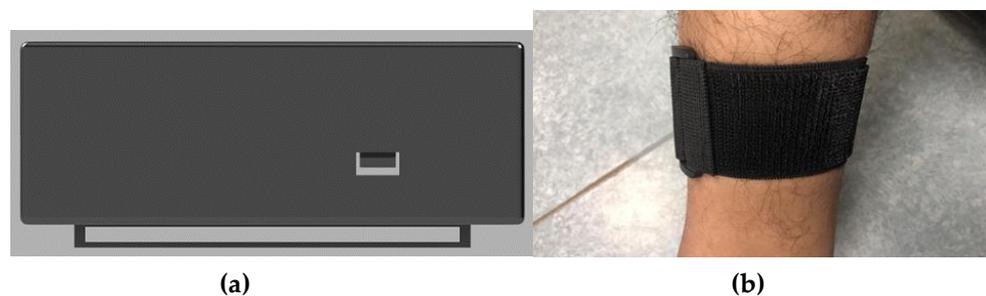

Figure 10: (a) Bottom Side of the Box with Space to Attach the VELCRO Strap; (b) A Demonstration of "How the (flexible) VELCRO will Attach the Box to the leg."

## 3. Results

This section provides the results of the various blocks mentioned in section 2.

### 3.1 Pressure Sensor Selection

#### 3.1.1. Velostat Characterization

The used light loads are in the range of 0.172 to 5.250 Kg, it can be taken as a reference to heavy loads. One test was done with sampling time of 7 seconds and another test was implemented with approximately 1-2 seconds. Both plotted graphs gave an expected result as shown in Figure 11 below. As the loads, increase the gravitational force increase causing the resistance to decrease in noticeable way. However, for the 7-second sampling time the hysteresis looks smaller while for the 1 second sampling the hysteresis effect is noticeably greater that illustrates that 1 second is not enough for the sheet to return to it is original value and there is a clear lagging in the response. For the intended purpose, a stable and fast response sensor is required to detect the fast changes in the movement of feet.

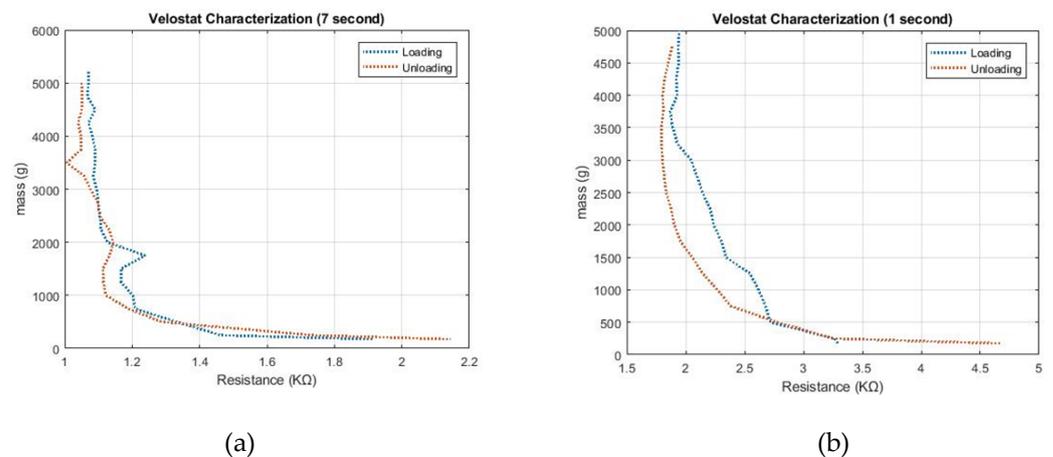

(a)            (b)

Figure 11: Resistance versus Weight for (a) 7 seconds (b) 1 second

### 3.1.2 FSR Characterization

The 500g weights were placed one by one every 2 to 3 seconds until reaching 10000g and the resistance was measured by a multimeter with every change in weight applied. Figure 12 below illustrate the relation between the resistance in logarithmic scale and the weight applied. It can be seen that the resistance decreases by increasing the load and by unloading the resistance increases again; however, it shows hysteresis since for the same weight the resistances when loading and unloading are not similar.

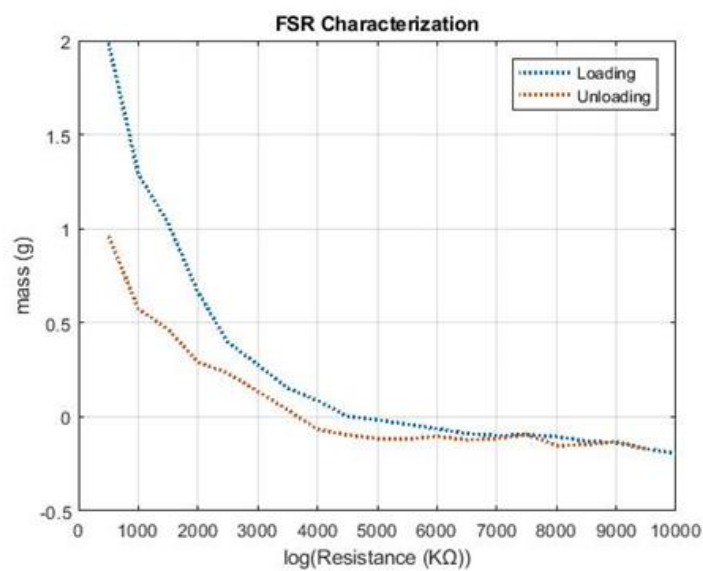

Figure 12: Characterization of the FSR

For initial characterization, 9 FSR sensors were placed on an insole not 16. However, they were placed at the targeted pressure points according to the literature review [2, 20, 22, 23]. Moreover, without applying pressure the resistance of all the sensors were infinity; therefore, all the multimeter displayed open loop (OL) while when a subject stood on the insole and applied full pressure the resistances decreases as in Figure 13 (b).

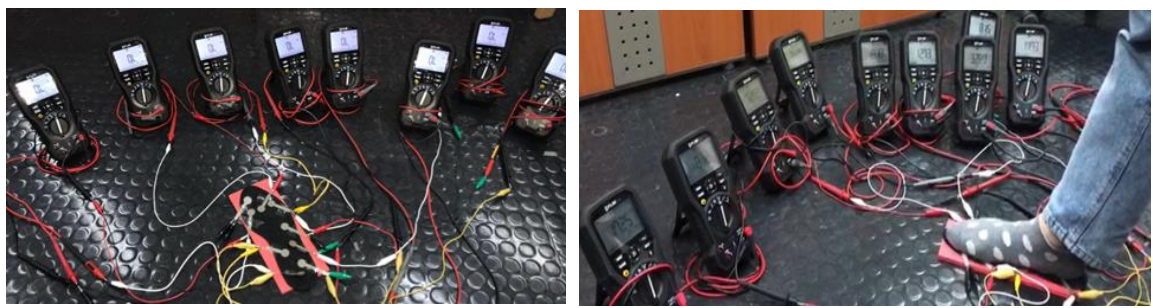

(a)          (b)

Figure 13: FSR Testing (a) without (b) with Applying Load

### 3.1.3. Piezoelectric Characterization

A mobile phone on vibration mode was placed on the sensor and called to observe the voltage when vibration occurs. Figures 14 are two responses for two different phones with two different vibration amounts was applied on the sensors to provide constant amount of vibration. The dynamic voltage peaks are almost the same when same amount of vibration is applied.

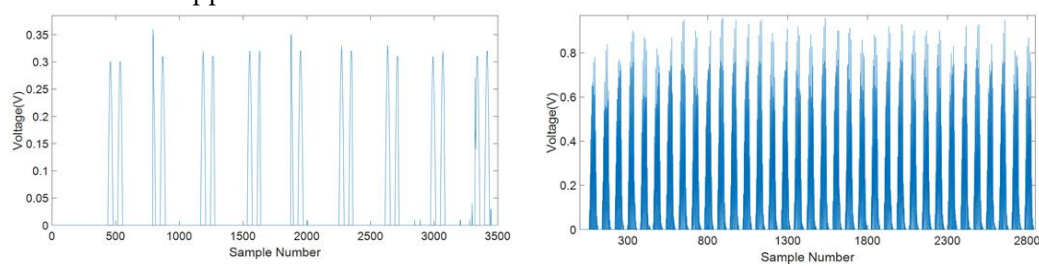

(a)          (b)

Figure 14: Dynamic Voltage Output of (a) Sample 1 (b) Sample 2

Due to cost comparison, response and range of the sensors, FSR was selected as the final sensor for the insole , it was also confirmed in previous studies done by the authors [20].

### 3.2 Temperature Sensor Selection

Based on the criteria discussed in Section 2, the temperature sensor module shown in Figure 15 was chosen for the insole design. The sensor is an NTC (Negative Temperature Coefficient) 10k Thermistor, which means that the resistance will decrease as the temperature increases (common type, another one is PTC), and the value of the resistance is 10kΩ at room temperature (25°C). The sensor did not need any calibration since the calibration data was provided by the manufacturer.

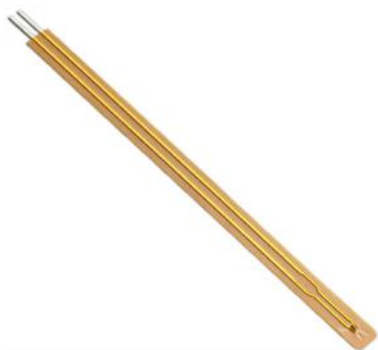

Figure 15: 10k NTC Thermistor Module from Littlefuse [40]

### 3.3 FinalDesigned Foot Sole

Before inserting the FSRs into the insole, critical points around the feet were checked based on similar research and fitting with a person's feet. Here only the "Active Area" of the FSRs were placed on the upper surface of the insole, and the tail was dipped inside through a hole and sewed on the bottom side in order to make them stable. The PCB design is deeply connected to the insole design and the connector used between the insole and the PCB.

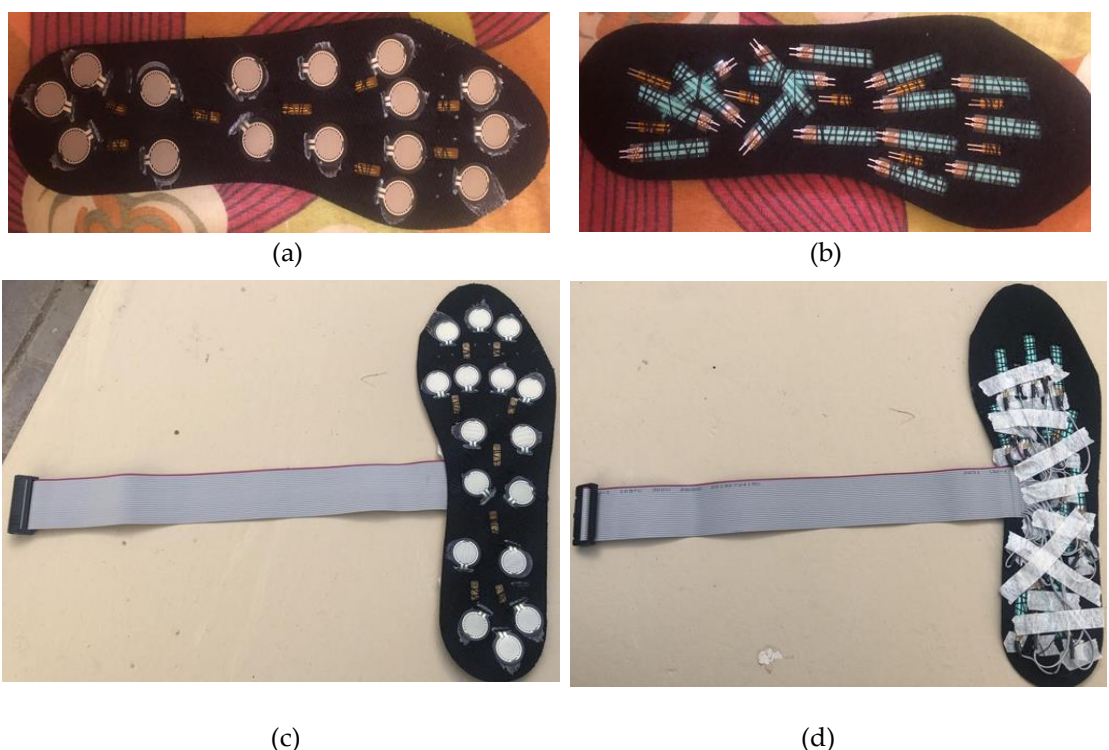

(a)  (b)

(c)  (d)

Figure 16: (a) FSRs and Temperature Sensors Glued to the Front End of the Insole; (b) Bottom Side of the Insole; (c) The Complete Insole after being Soldered to the Ribbon Cable (Top Side); (d) Bottom Side of the Completed Insole

After the FSRs were soldered, all of them were carefully and thoroughly tested for continuity in the cable by short circuit test, open circuit test for the FSR (FSRs are open circuit without any force applied on it) and variable external pressure was applied to check for overall response. Eight flexible temperature sensors were embedded into each insole, alongside the sixteen FSRs. After adding the temperature sensors, the total number of sensors per insole was 24, and one pin is needed for the ground (or the VCC, depends on the PCB design). So, a ribbon cable with 26 pins (13 pins in each row was chosen), and the PCB model (3rd and 4th) was designed to be compatible with it. Then the ribbons had been soldered to the insole after the insole was ready. The final sole designed can be seen in Figure 16.

**3.4 PCB Design**

For minimizing and improving the PCB had been a significant goal achieved in this project by identifying the efficient and possible ways of improvement. Recognizing the process of designing a PCB layout is a major key for optimizing the PCB and its performance. The following steps were followed while designing the PCB:

Before designing the PCB, a schematic is required to draw to verify the electrical connections of the PCB. It acts as a backbone for the PCB Layout. Both DipTrace and Altium software was utilized to design the PCB (DipTrace was used mainly due to the availability of Free Student License). In DipTrace (also in Altium), the schematic can be directly converted to a PCB Layout, which can be formatted afterward. It lessened the work to a great extent. In Figure 18, final PCB (double layered) has been showed. Using Surface Mounted Technology (SMT) components (due to their smaller size and compact shape) to downsize the PCB.

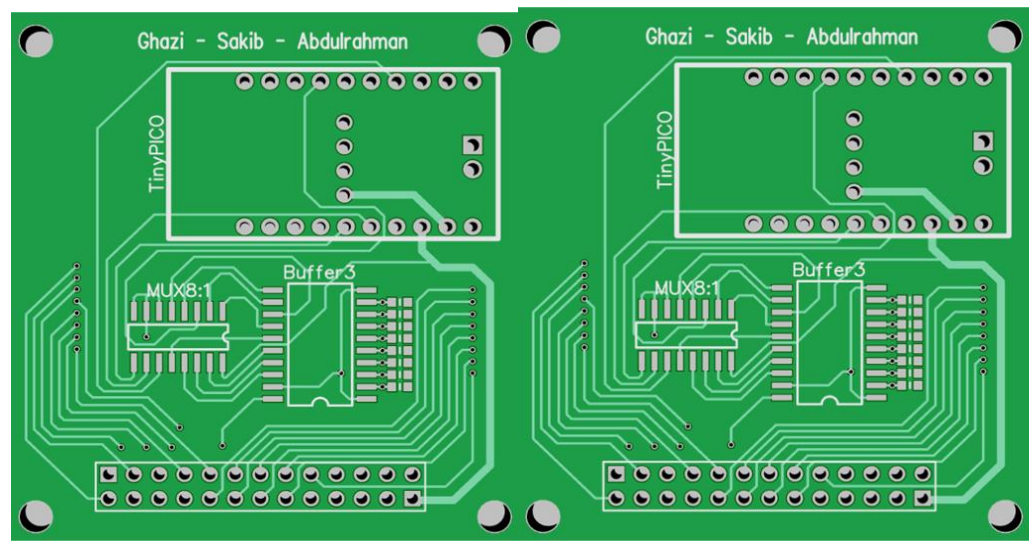

(a) (b)

Figure 17: (a) Top Layer, (b) Bottom Layer of the PCB Design Model (3D View)

PCB Soldering and Connecting to the Insole: One of the PCBs was soldered to all its components, as shown in Figure 18 (a-c). After the PCB was soldered, the smart insole was attached to the PCB through the ribbon cable (Figure 18 (d)). The Ribbon cable, which was chosen carefully beforehand, fitted tightly to the hardware, thus making a rigid connection between the insole and the hardware. The PCB components which were soldered to the printed PCB board are TinyPico, Inertial Measurement Unit (IMU), Male headers for attaching the ribbon cable, JST Connector in the bottom side to connect the battery; Surface Mounted (SMD) resistors and capacitors, mux, buffer, and other chips.

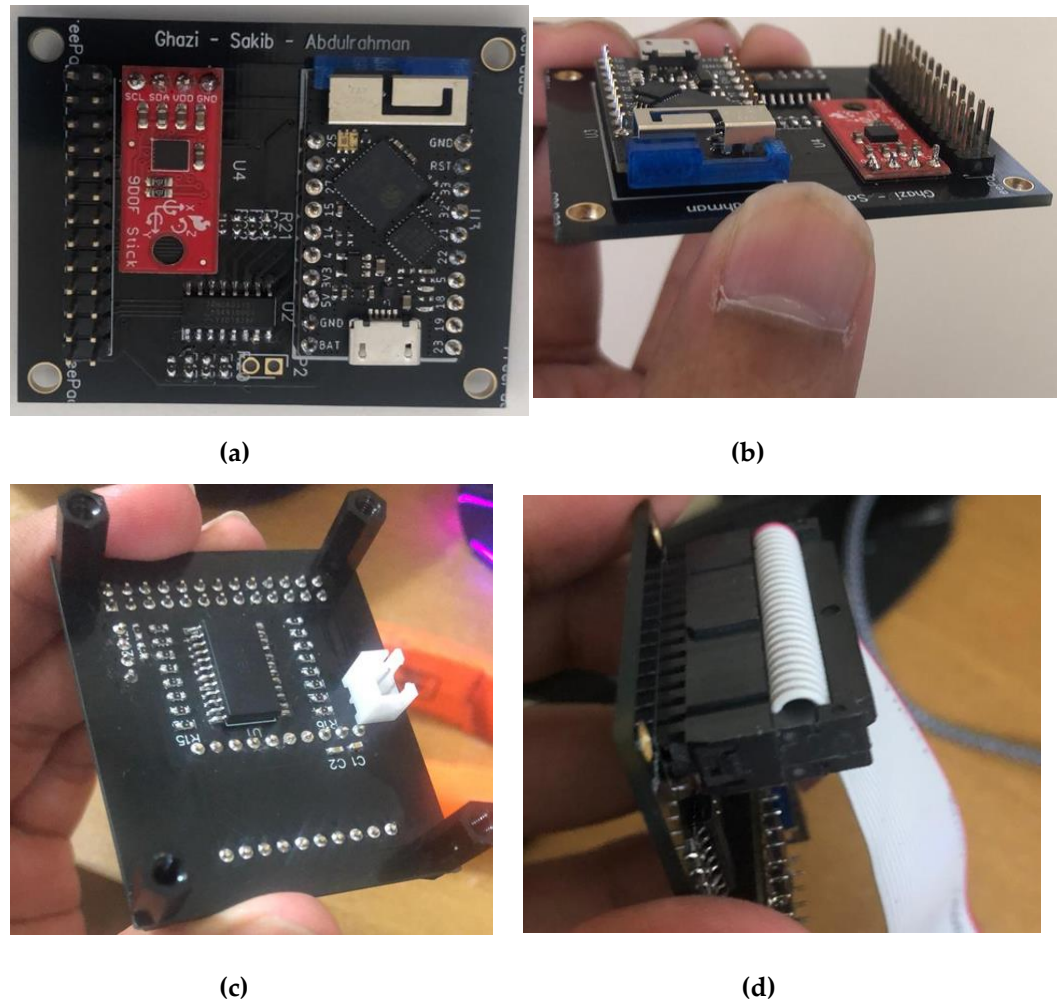

Figure 18: PCB Soldered (a) Top View; (b) Side View; (c) Bottom View; (d) Attached to the via Ribbon Cable

### 3.5 Bluetooth communication

Bluetooth.com provides a "Bluetooth Range Estimator" service where some common parameters are used, along with some assumptions to estimate the Bluetooth range of a device. From the esp32 datasheet, the values of these parameters were recorded as follows:

- Maximum Data Transfer Rate: BLE/BT (1 MHz)
- Receiver Sensitivity: -97 dBm [38]
- Transmit Power: 0 dBm [41]
- Path Loss: Office (Qatar University Lab)
- Transmitter Antenna Gain: 2 dBi [42]
- Receiver Antenna Gain: 2 dBi [42]

Inputting these values provided a nominal range of 17 to 23 meters between two BLE devices, maintaining a P2P (peer-to-peer) connection. The range was tested in a lab at Qatar University (QU), and it was approximately 15 meters, which is within the range constraint of the smart insole (considering the noisy environment of the lab). If the smart insole can work within a range of 15-20 meters, that is more than enough since the gait cycle analysis of a patient can smoothly be done within this range. But nevertheless, performing the experiment outdoor will provide around 4x range.

### 3.6 Generated Temperature, Pressure Maps and corresponding Gait cycles

The data were collected from 12 subjects. Subjects characteristics is shown in Table 2.

Table 2: Details of subjects whose readings were taken during the test

| Number of Subjects | Age (Year) | Weight (kg) | Height (cm) | BMI | Genders |
|---|---|---|---|---|---|
| 12 | 20-59 | 52-125 | 153-185 | 18-36.5 | Female and Male |

All subjects were asked to walk a 10 m walkway with self-selected cadence for 6 times and data acquired at 40Hz frequency from 16 FSRs that were converted into force and represented in columns. The summation of each row was calculated to obtain the whole gait cycle performed by the subject at one trial. Figure 19 illustrates the whole gait cycle of a subject which shows that almost each gait cycle is represented by two peaks with the second peak much higher than the first and the duration of each gait cycle is not constant through the whole trial.

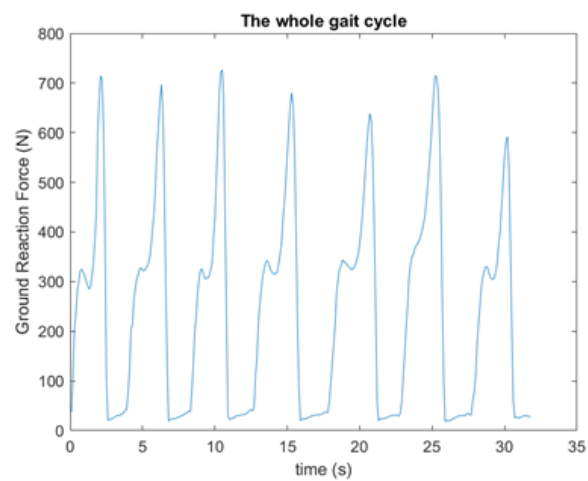

Figure 19: Gait cycle for one trial

Each gait cycle was segmented from the whole gait cycle, resampled to $2^9=512$ samples for all the gait cycles in all trials for all subject and normalized as seen in Figures 20 and 21, respectively, to facilitate the comparison between all the gait cycles. Moreover, the mean of all the resampled gait cycles is plotted along with the minimum and maximum to show the standard deviation of the gait cycles to the mean as Figure 22.

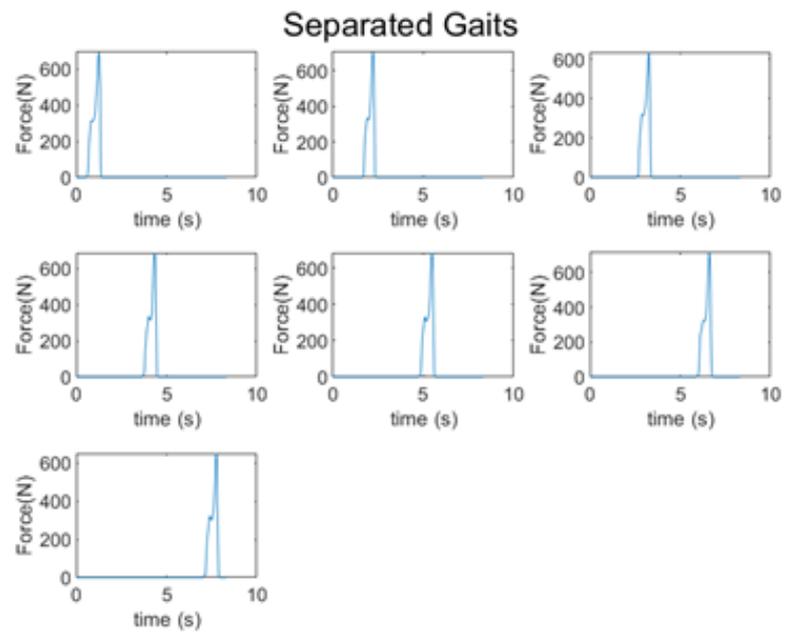

Figure 20. segmentation of the whole gait cycle

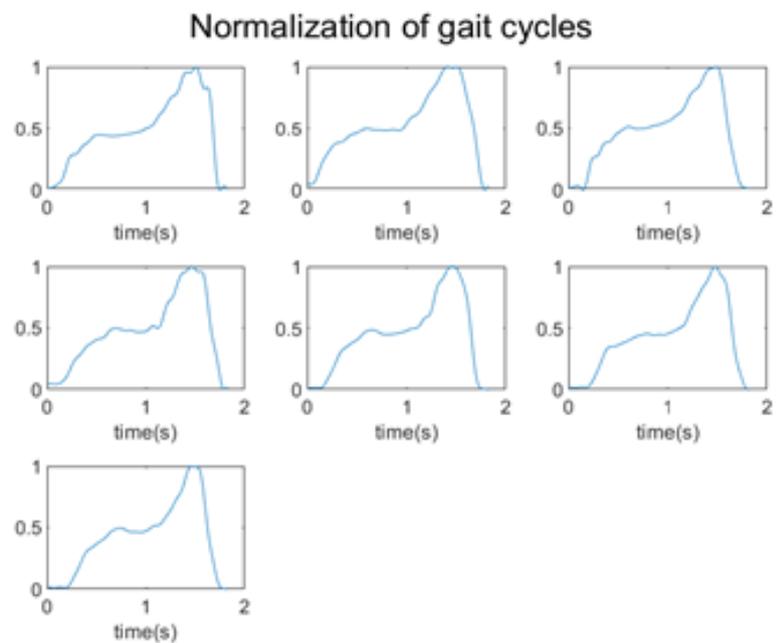

Figure 21. Normalization of the segmented gaits

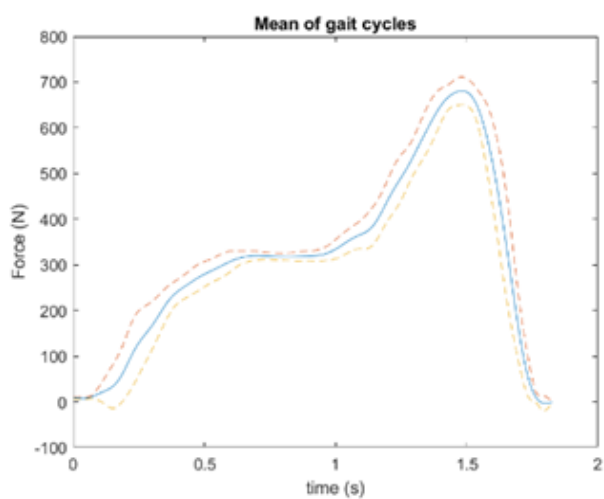

Figure 22.   Mean of all gait cycles

The readings taken from the FSR sensors could be presented as a visual pressure map that show the areas exposed to pressure during the gait cycle. The idea of the pressure map was implemented first by designing a foot template with specified location of sensors, after that the force values were filled in the template. By using MATLAB software, the data were localized in the template then interpolated to estimate the value of pressure on the areas that does not have a direct sensor attach to it. The pressure map could illustrate the stages of the gait cycle that is divided to stance and swing phases, the former is mainly applying the whole-body weight to the ground and it occupy 60% of the total gait cycle it consists of Heel strike, foot-flat, midstance and heel-off stages. The latter begins when the toes left the ground and ends when the heel strikes it, it consists of toe-off, mid-swing and terminal swing (heel strike) stages.

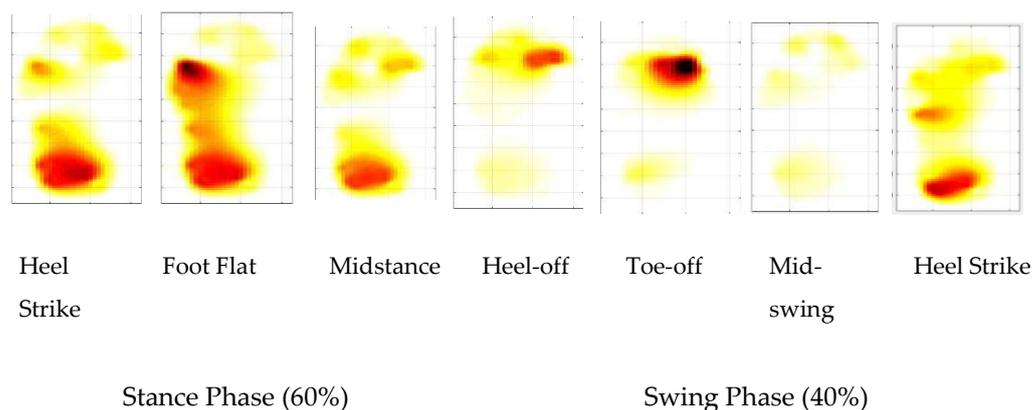

| Heel Strike | Foot Flat | Midstance | Heel-off | Toe-off | Mid-swing | Heel Strike |

Stance Phase (60%)                              Swing Phase (40%)

Figure 23.   Pressure Map during different phases of the Gait Cycle

Figure 24 shows the temperature maps from the insole, when the user was asked to stand still , as the standing temperature would be of interest in early detection of foot complication[23] .

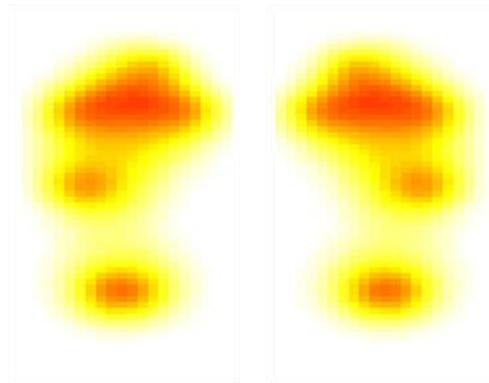

Figure 24.   Temperature Map during the stance position

## 4. Discussion and Analysis

### 4.1 Calibration of FSR pressure sensor

The sensors were tested for their response of dynamic pressure since during locomotion, the person's feet will apply dynamic pressure (variable over instant) on the sensors placed in the foot sole, and among the three types of sensor, FSRs were found to be the most suitable for placing on the insole. Even though FSRs were selected as they were found to be the most suitable to be used in an insole after testing their dynamic response, using FSRs alone (by connecting FSRs in series from the power supply to the ground) will not provide any useful output since FSR is just a "variable resistor which is reactive to force applied on it." FSR based applications (including this one) require reading the response of the FSR (change in resistance; that's what makes them Force Sensitive Resistors) with respect to the force applied on it using the Analog to Digital Converter (ADC) of the microcontroller. For this purpose, a voltage divider circuit needs to be implemented for each FSR, as shown in Figure 25 (a). The equation for the voltage divider is as follows:

$$V_{out} = V_{cc} * \left(\frac{R_{ext}}{R_{ext} + R_{FSR}}\right) \Rightarrow R_{FSR} = \frac{V_{cc} * R_{ext}}{V_{out}} - R_{ext} = \frac{R_{ext}(V_{cc} - V_{out})}{V_{out}} \dots (1)$$

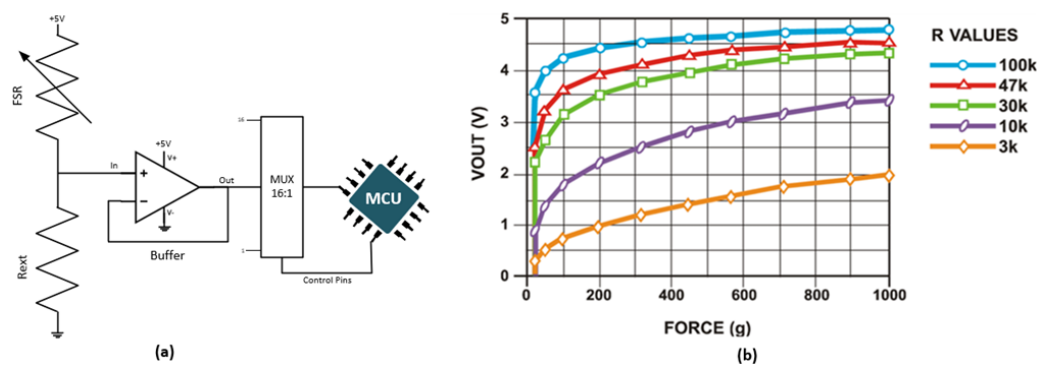

Figure 25: (a) Voltage Divider Circuit to take FSR Measurements, (b) Voltage Output Response for Different Values of Pull-Down (or Pull-Up) Resistor with Response to Force Applied on the FSR [43]

In normal conditions, without any force applied on it, FSRs show an infinite resistance, which decreases gradually as the force increases, i.e., they have an inverse relationship, but the relationship is not linear. Focusing on the voltage divider circuit shown in Figure 25 (a), it can be understood that using a 10kΩ Rext will provide the best (and slowest) response. It is because the voltage divider circuit works robustly when both resistances are compatible in terms of magnitude, and the input voltage gets divided (almost) evenly. But the sensing performance deteriorates (i.e., readings in ADC or current flow) abruptly or changes quickly to saturation if the ratio of their magnitude is high (e.g., one is mega-ohm, and another is ohm). Now, according to the plot in Figure 25, as provided by the Interlink Electronics [43], FSR resistance is approximately 10kΩ for an applied force of 100g over the surface of the FSR. But this force is much lower than the mean peak force applied by an adult human in various regions under the feet. According to the outcomes of the Ph.D. thesis of Jerman Perttunan, the peak pressure under different areas of the feet of an adult maintain 60% of the Gait Cycle in the stance phase has the pattern shown in Figure 26 [44]. So, the maximum pressure any FSR will face varies in each region under the feet. According to the plot in Figure 26, peak plantar pressure in various regions under the feet of an average adult human varies from 80kPa to around 600kPa.

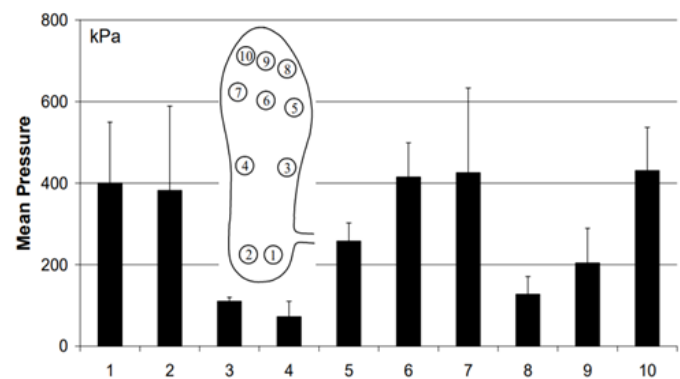

Figure 26: Mean and SD of Planter Peak Pressure (in kPa) in various foot regions [44]

The information about the active sensing area of the FSRs used can be found in their respective datasheets[45]. The active sensing region for FSR-402 is a circular region with a radius of 6.35 mm. That means, area of the sensing region, $A=\pi R^2 \cong$ **126.677 mm²**. On the other hand, we know 1 Pascal (Pa) $\equiv$ 1 $Nm^{-2}$, i.e., equivalent to 1N of force on an area of a one-meter square. So, the force applied on the FSRs embedded to the insoles varies from $\frac{80000*126.677}{1000*1000} = \mathbf{10N}$ to $\frac{600000*126.677}{1000*1000} = \mathbf{76N}$. But considering the nominal value of gravitation, g = 9.81$ms^{-2}$. We can convert this force or weight into the mass. So, the mass varies from **1.03kg** to **7.75kg**. **4.2 Calibration of NTC sensor**

Based on the manufacturer data, sensor response was plotted for the measuring range (20°C to 50°C), and the equation (Resistance vs. Temperature relation) was found in MS Excel using Excel's built-in Exponential Fitting (figure 27). Three types of fitting viz. Linear, Polynomial, and Exponential Fitting were tried for this purpose. Polynomial

Fitting and Exponential Fitting gave a similar level of accuracy and preciseness in fitting with an R2 value of 0.996, which is high (much higher than the result found through linear fitting). But the polynomial fitting w snot chosen since we need the value of the temperature based on the resistance reading from the sensor (i.e., we need x in terms of y), but the opposite is given. This equation was used in the code to transform the analog reading of the sensor directly to a temperature value.

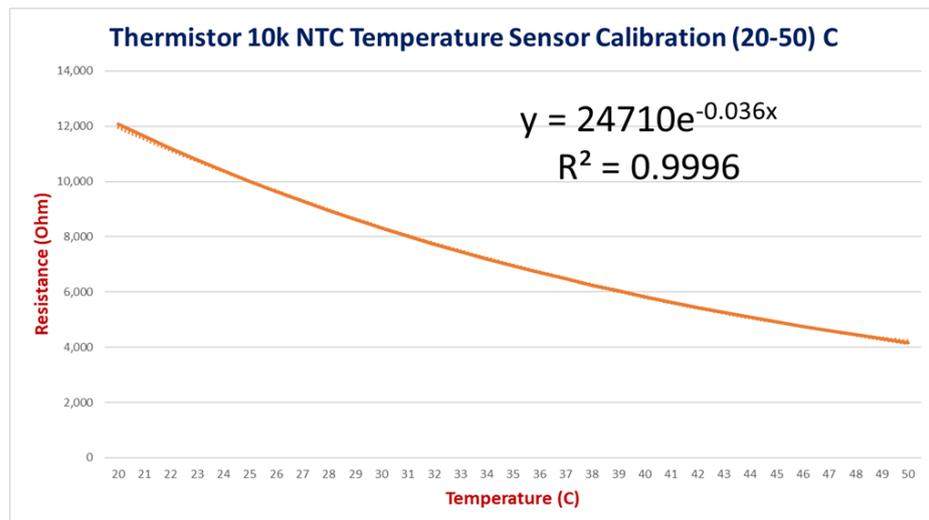

Figure 27: Thermistor Resistance vs. Temperature Plot and Exponentially Fitted

The derivation of the equation for Temperature vs. Analog Reading is as follows:

$$V_{out} = \frac{V_{cc} * R_{ext}}{R_{ext} + R_{Temp}} = \frac{5 * 10000}{10000 + R_{Temp}} = \frac{50000}{10000 + R_{Temp}} \quad [\text{Since}, V_{cc} = +5V, \text{and } R_{ext} = 10k\Omega] --(2)$$

Here, $R_{Temp} \equiv R_{FSR}$ since both follow the voltage divider rule. Now, since, esp32 has a 12-bit ADC, the number of ADC levels is $2^{12} = 4096$, i.e., the ADC value ranges from 0 to 4095. 0 to 4095 value is mapped into 0 to 5V voltage supply, linearly. So, the relation between the output voltage and the ADC readings can be represented by, $\frac{V_{out}}{5} = \frac{ADC\ Reading}{4095} \Rightarrow V_{out} = \frac{A}{819}$ [Let, ADC Reading $\equiv A$]

So, we have, from (2),

$$\Rightarrow \frac{A}{819} = \frac{50000}{10000 + R_{Temp}} \Rightarrow A = \frac{40950000}{10000 + R_{Temp}} \Rightarrow R_{Temp} = \frac{40950000 - 10000 * A}{A} ---(3)$$

Now, from the calibration equation from Figure 27, we have,

$$y = 24710 * e^{-0.036x} \equiv R_{Temp} = 24710 * e^{-0.036*T} \quad [\text{Let}, 'T'\ \text{denotes\ Temperature}]$$

$$\Rightarrow \ln(R_{Temp}) = \ln(24710 * e^{-0.036*T}) = \ln(24710) + \ln(e^{-0.036*T}) \quad [\text{Taking\ Natural\ Logarithm}]$$

$$\Rightarrow \ln(R_{Temp}) = 10.115 - 0.036T \Rightarrow T = \frac{\ln(R_{Temp}) - 10.115}{-0.036} \ldots (4)$$

From equation (3) and (4), we get,

$$\text{Temperature}, T \approx \frac{\ln\left(\frac{40950000 - 10000 * A}{A}\right) - 10.115}{-0.036} \quad (6)$$

Where we can get the value of 'A' or the analog reading from the sensor output. Note that this formula will only hold when the external resistance used is 10kΩ.

**4.3 Multiplexing FSR and Temperature Sensor Data**

There are 16 FSRs and 8 Temperature Sensors in each insole. Readings from them need to be multiplexed and represented in a single string before transferring the data wirelessly to the central device (connected to a PC) from the peripheral (connected to the insole). This is because 16 FSRs will produce 16 analog values per sample and eight analog values from the eight temperature sensors. So, the MCU will need 24 onboard Analog to Digital Converter or ADCs if the data is not multiplexed or combined. For this purpose, a 16:1 MUX (multiplexer) for the FSR and an 8:1 MUX for the temperature sensors are used in the circuit, which would require only 2 ADC instead of 24. The total packet size after combining the two MUX data became around 120 bytes. After adding the header information like Server and Characteristic UUIDs, the most any user could for data transfer was 20 bytes only. So, the packets were sent in small chunks; otherwise, part of the packets was being expurgated. Bluetooth v4.2 got rid of this issue by extending the packet size to 255 bytes. Since v4.2 is supported by esp32, this problem was solved. In order to calculate the maximum packet size, all FSR and Temperature data were sent to the central device wirelessly via Bluetooth, and the received data was shown in Arduino IDE's Serial Monitor represented as strings (shown in Figure 28).

```
-0.13    -0.24     0.96   -0.64  -15.71   -5.46   -0.09     0.48   -0.54  -14.02    7.28   -0.44
   0    0    0    0    0    0    0    0    0    0    0    0    0    0    0    0

-0.17    -0.22     0.91   -8.35  -13.33   -5.36   -0.09     0.47   -0.54  -13.86   10.61   -4.47
   0    0    0    0    0    0    0    0    0    0    0    0    0    0    0    0

-0.03    -0.08     0.75  -15.51   35.33   -0.12   -0.06     0.46   -0.53   -5.95    2.37   -0.64
   0    0    0    0    0    0    0    0    0    0    0    0    0    0    0    0

-0.04    -0.20     1.01   -4.39    7.78    8.09   -0.05     0.45   -0.53  -11.43    2.18   -0.65
   0    0    0    0    0    0    0    0    0    0    0    0    0    0    0    0
```

Figure 28. FSR and Temperature Data String (Screenshoted from Arduino IDE's Serial Monitor)

Then the size of that string was calculated in terms of bytes using an online string size calculator, as shown in Figure 29.

```
UTF-8 string length & byte counter

-0.05    -0.16    1.00    -29.35    -31.72    5.52    -0.06    0.44    -0.53
-9.18    2.84    -4.03
4095 4095 4095 4095 4095 4095 4095 4095 4095 4095 4095 4095 4095 4095
4095 4095
```

That's **173 characters**, totaling **173 bytes**. #

Figure 29. Online Packet Size Calculator (Approximation) [46]

Here, from the online packet size calculator, our string or packet size containing all sensor data is 118 bytes. The FSR Data can go up to 4 bytes for each of 16 FSR reading and 15 bytes for space between adjacent readings equal to $4*16+16 = \mathbf{80\ bytes}$. And, the Temperature Sensor Data can go up to 4 bytes for each of 8 sensors reading and 7 bytes for space between adjacent readings equal to $4*8+7 = \mathbf{39\ bytes}$.

The highest possible packet size = (80+ 39) bytes = 119 bytes ≡(119*8) bits=952 bits, which is well inside the maximum allowable packet size of 251 bytes (255 bytes in total, 4 bytes for the header) for Bluetooth v4.2 supported by esp32. Now, the maximum data transfer rate allowed by BLE v4.2 is 1 Mbps (Megabit Per Second), as discussed earlier. But this speed changes as the distance or the packet size increases and vice-versa.

**4.4 . Power Consumption Test for the MCU in Bluetooth LE Communication**

Even though Bluetooth LE has been designed as a low power consuming Communication Protocol, the effectiveness of BLE in this solution needed to be checked practically. Bluetooth LE has a Current Draw Cycle like the one in Figure 30. Here BLE remains in the "Sleeping" mode most of the time for its operation while it only "Wakes Up" during data transmission.

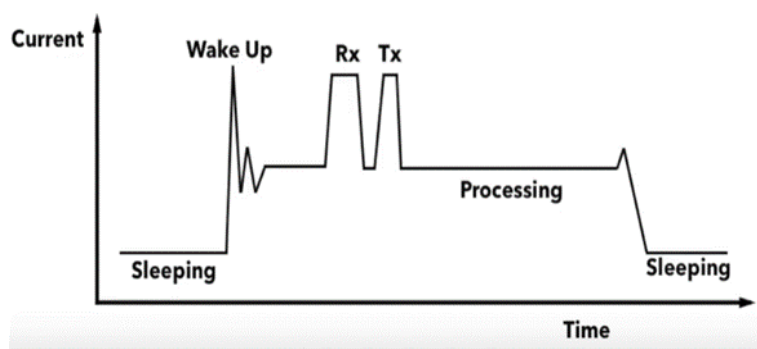

Figure 30: Current Draw Cycle of Bluetooth Low Energy or BLE [46]

For BLE, the most amount of power is consumed during data transmission. This is the fundamental difference between traditional Bluetooth and Bluetooth LE (Previously Bluetooth Smart). But in this case, the patient will be wearing the device during testing, and data transmission will occur continuously during this time. So, the device will not have the chance to enter into any power-saving mode. Low power modes play a crucial role when the device needs to be turned on for a long time and requires sending data occasionally. The exact values of current draw (i.e., power consumption) are more device-dependent than the communication protocol itself. Moreover, the other components of the device require power as well. So, the ideal power consumption provided by the Bluetooth authority for BLE is not the only power the battery requires to supply during data acquisition. Here, Bluetooth (BT), or Wi-Fi cannot function during any low power mode. Now, based on this, the power consumption of the esp32 module during BLE data transfer was checked using a Digital Multimeter, as shown in Figure 31. The USB represents the current draw (A), Voltage (V) and Power Consumption (W) per unit time, and the energy consumption in Watt-hour (Wh). Both the current draw per unit time and energy consumption were studied in order to understand the power consumption of the device during BLE operation. The USB Tester has a timer in it. Data transfer via BLE was performed for 1 hour in order to record the total energy consumption. It was observed that the current drawn was varying over time, roughly within 0.05A to 0.14A. But the total energy consumption recorded was about **0.4 Wh,** i.e., in one hour, the device consumed around 0.4W power. Mentionable that the voltage supplied for the esp32 was approximately 5V.

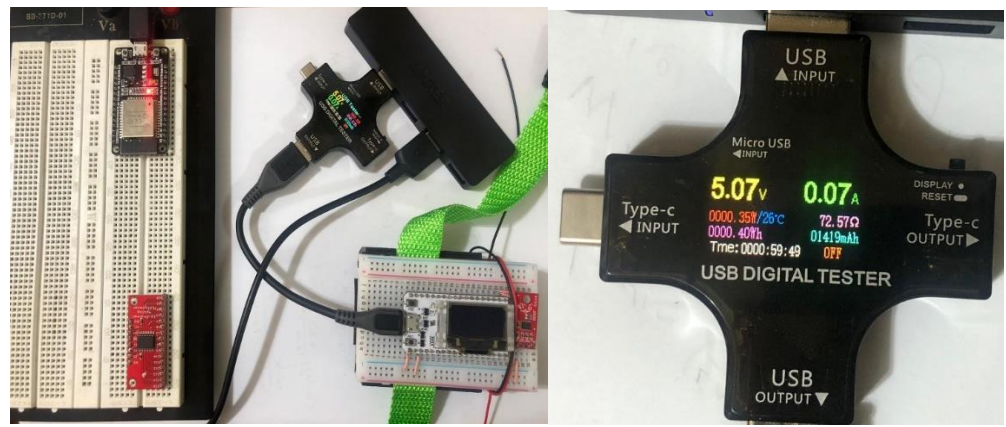

Figure 31: Power Consumption Test for the BLE Peripheral Device Over Full IMU and MUX Data Transfer

For performing a simple calculation for battery life, let the battery capacity be 1300 mAh (since Li-ion or LiPo batteries are planned to be used for their rechargeability, their service life is normally measured in terms of mAh). Li-ion or LiPo batteries supply 3.7V. So, the energy rating of the battery will be around **4.81Wh**. Then the battery can safely run the circuit for $\left(\frac{4.81}{0.4}\right)$ Hours $\cong$ **12 hours**. So, if someone wants to use the device half an hour a day, it will take him/her about **24 days** to deplete the battery from 100% without recharging it.

5. **Conclusion**

In this study, the authors have proposed and designed low-cost calibration setups for calibrating temperature and pressur sensors. The FSR-based smart insole was able to acquire high quality vGRF for different gait cycles. The NTC thermistor was used for plantar temperature map generation. IN addition the paper provides details of the communication protocol and microcontrollers that can be used for such application. The system also provides details of how the foot pressure and temperature data from the subjects using the sensors can be transferred wirelessly using a low-power consuming communication protocol to a central device where the data will be recorded. The investigation can help in developing low-cost, feasible and portable foot monitoring system for patients by facilitating real-time, home monitoring of foot condition using Gait Cycle or Foot Pressure patterns and temperature heterogeneity between two feet. . The proposed system will work in real-time can be used for remote health monitoring at the convenience of the user which when incooporated with Artificial Intelligent Solution can help in early detection of foot complication.

**References**

[1] M. E. Chowdhury, A. Khandakar, Y. Qiblawey, M. B. I. Reaz, M. T. Islam, and F. Touati, "Machine learning in wearable biomedical systems," in *Sports Science and Human Health-Different Approaches*, ed: IntechOpen, 2020.

[2] A. Khandakar, M. E. Chowdhury, M. B. I. Reaz, S. H. M. Ali, M. A. Hasan, S. Kiranyaz, *et al.*, "A machine learning model for early detection of diabetic foot using thermogram images," *Computers in Biology and Medicine,* vol. 137 p. 104838(2021)

[3] T. Rahman, F. A. Al-Ishaq, F. S. Al-Mohannadi, R. S. Mubarak, M. H. Al-Hitmi, K. R. Islam, *et al.*, "Mortality Prediction Utilizing Blood Biomarkers to Predict the Severity of COVID-19 Using Machine Learning Technique," *Diagnostics,* vol. 11 (9), p. 1582(2021)

[4] A. Al Omar, A. K. Jamil, A. Khandakar, A. R. Uzzal, R. Bosri, N. Mansoor, *et al.*, "A Transparent and Privacy-Preserving Healthcare Platform With Novel Smart Contract for Smart Cities," *IEEE Access,* vol. 9 pp. 90738-90749(2021)

[5] A. Khandakar, M. E. Chowdhury, M. B. I. Reaz, S. H. M. Ali, M. A. Hasan, S. Kiranyaz, *et al.*, "A Machine Learning Model for Early Detection of Diabetic Foot using Thermogram Images," *arXiv preprint arXiv:2106.14207,* 2021)

[6] C. Tavares, F. Leite, M. D. F. Domingues, T. Paixão, N. Alberto, A. Ramos, *et al.*, "Optically Instrumented Insole for Gait Plantar and Shear Force Monitoring," *IEEE Access,* vol. 9 pp. 132480-132490(2021)

[7] S. Bus, S. Van Netten, L. Lavery, M. Monteiro-Soares, A. Rasmussen, Y. Jubiz, *et al.*, "IWGDF guidance on the prevention of foot ulcers in at-risk patients with diabetes," *Diabetes/metabolism research and reviews,* vol. 32 (S1), pp. 16-24(2016)

[8] A. M. Reyzelman, K. Koelewyn, M. Murphy, X. Shen, E. Yu, R. Pillai, *et al.*, "Continuous temperature-monitoring socks for home use in patients with diabetes: observational study," *Journal of medical Internet research,* vol. 20 (12), p. e12460(2018)

[9] R. G. Frykberg, I. L. Gordon, A. M. Reyzelman, S. M. Cazzell, R. H. Fitzgerald, G. M. Rothenberg, *et al.*, "Feasibility and efficacy of a smart mat technology to predict development of diabetic plantar ulcers," *Diabetes Care,* vol. 40 (7), pp. 973-980(2017)

[10] F. N. Inagaki Nagase, "The impact of diabetic foot problems on health-related quality of life of people with diabetes," 2017)

[11] P. A. Crisologo and L. A. Lavery, "Remote home monitoring to identify and prevent diabetic foot ulceration," *Annals of translational medicine,* vol. 5 (21), 2017)

[12] M. El-Nahas, S. El-Shazly, F. El-Gamel, M. Motawea, F. Kyrillos, and H. Idrees, "Relationship between skin temperature monitoring with Smart Socks and plantar pressure distribution: A pilot study," *Journal of wound care,* vol. 27 (8), pp. 536-541(2018)

[13] K. Deschamps, G. A. Matricali, P. Roosen, K. Desloovere, H. Bruyninckx, P. Spaepen, *et al.*, "Classification of forefoot plantar pressure distribution in persons with diabetes: a novel perspective for the mechanical management of diabetic foot?," *PLoS One,* vol. 8 (11), p. e79924(2013)

[14] J. W. Albers and R. Jacobson, "Decompression nerve surgery for diabetic neuropathy: a structured review of published clinical trials," *Diabetes, metabolic syndrome and obesity: targets and therapy,* vol. 11 p. 493(2018)

[15] N. C. Silva, H. A. Castro, L. C. Carvalho, É. C. Chaves, L. O. Ruela, and D. H. Iunes, "Reliability of infrared thermography images in the analysis of the plantar surface temperature in diabetes mellitus," *Journal of chiropractic medicine,* vol. 17 (1), pp. 30-35(2018)

[16] B. Lahiri, S. Bagavathiappan, B. Raj, and J. Philip, "Infrared thermography for detection of diabetic neuropathy and vascular disorder," in *Application of Infrared to Biomedical Sciences*, ed: Springer, 2017, pp. 217-247.

[17] W. L. Schneider and M. Severn, "Prevention of plantar ulcers in people with diabetic peripheral neuropathy using pressure-sensing shoe insoles," 2018)

[18] B. Najafi, H. Mohseni, G. S. Grewal, T. K. Talal, R. A. Menzies, and D. G. Armstrong, "An optical-fiber-based smart textile (smart socks) to manage biomechanical risk factors associated with diabetic foot amputation," *Journal of diabetes science and technology,* vol. 11 (4), pp. 668-677(2017)

[19] A. Oks, A. Katashev, M. Zadinans, M. Rancans, and J. Litvak, "Development of smart sock system for gate analysis and foot pressure control," in *XIV Mediterranean Conference on Medical and Biological Engineering and Computing 2016*, 2016, pp. 472-475.

[20] A. M. Tahir, M. E. Chowdhury, A. Khandakar, S. Al-Hamouz, M. Abdalla, S. Awadallah, *et al.*, "A systematic approach to the design and characterization of a smart insole for detecting vertical ground reaction force (vGRF) in gait analysis," *Sensors,* vol. 20 (4), p. 957(2020)

[21] A. H. Abdul Razak, A. Zayegh, R. K. Begg, and Y. Wahab, "Foot plantar pressure measurement system: A review," *Sensors,* vol. 12 (7), pp. 9884-9912(2012)

[22] A. Khandakar, M. E. Chowdhury, M. B. I. Reaz, S. H. M. Ali, T. O. Abbas, T. Alam, *et al.*, "Thermal Change Index-Based Diabetic Foot Thermogram Image Classification Using Machine Learning Techniques," *Sensors,* vol. 22 (5), p. 1793(2022)

[23] U. Niemann, M. Spiliopoulou, T. Szczepanski, F. Samland, J. Grützner, D. Senk, *et al.*, "Comparative clustering of plantar pressure distributions in diabetics with polyneuropathy may be applied to reveal inappropriate biomechanical stress," *PLoS One,* vol. 11 (8), p. e0161326(2016)

[24] J. J. van Netten, J. G. van Baal, C. Liu, F. van Der Heijden, and S. A. Bus, "Infrared thermal imaging for automated detection of diabetic foot complications," ed: SAGE Publications Sage CA: Los Angeles, CA, 2013.

[25] *Thermal-Ribbon Flexible RTD and Thermocouple Temperature Sensors* [Online]. Available: https://www.mod-tronic.com/Rewind_Sensors/Minco_Thermal-Ribbon_Flexible_Sensors.html. [Accessed on: 01/02/2021]

[26] R. A. Nardin, P. M. Fogerson, R. Nie, and S. B. Rutkove, "Foot temperature in healthy individuals: effects of ambient temperature and age," *Journal of the American Podiatric Medical Association,* vol. 100 (4), pp. 258-264(2010)

[27] P. Mannion, "Comparing low-power wireless technologies (part 1)," *Digi-Key. Dec,* vol. 14 2017)

[28] G. Roth, "Bluetooth wireless technology," ed: Stanford University. Dostopno na: http://large. stanford. edu/courses/2012 …, 2013.

[29] (20019). *Maximizing BLE Range* [online]. Available: https://www.argenox.com/library/bluetooth-low-energy/maximizing-bluetooth-low-energy-ble-range/. [Accessed: 15- Nov- 2019]

[30] (2019). *Understanding Bluetooth Range | Bluetooth Technology Website* [Online]. Available: https://www.bluetooth.com/learn-about-bluetooth/key-attributes/range/#estimator. [Accessed: 15- Nov- 2019]

[31] *ESPRESSIF* [Online]. Available: https://www.espressif.com/. [Accessed on 01st January 2020]

[32] *Adafruit* [Online]. Available: https://www.adafruit.com/product/2821. [Accessed on 01st January 2020]

[33] *NORDIC SEMICONDUCTOR* [Online]. Available: https://www.nordicsemi.com/. [Accessed on 01st January 2021]

[34] *Arduino Nano BLE Sense* [Online]. Available: https://www.arduino.cc/en/Guide/NANO33BLESense. [Accessed on 01st January 2020]

[35] F. Touati, A. Khandakar, M. E. Chowdhury, S. Antonio Jr, C. K. Sorino, and K. Benhmed, "Photo-Voltaic (PV) Monitoring System, Performance Analysis and Power Prediction Models in Doha, Qatar," in *Renewable Energy*, ed: IntechOpen, 2020.

[36] *Beetle esp32* [Online]. Available: https://wiki.dfrobot.com/Beetle_ESP32_SKU_DFR0575#:~:text=Beetle%20ESP32%20is%20a%20simplified,size%20of%2035 mm%C3%9734mm. [Accessed on 01st January 2020]

[37] *TinyPICO* [Online]. Available: https://www.tinypico.com/. [Accessed on 01st January 2020]

[38] (2016, Accessed: 15- Nov- 2019). ESP-WROOM-02 PCB Design and Module Placement Guide. 1-7. Available: https://www.espressif.com/sites/default/files/documentation/esp-wroom-02_pcb_design_and_module_placement_guide_0.pdf

[39] *VELCRO* [Online]. Available: https://www.velcro.com/. [Accessed on 01st January 2020]

[40] (2022). [Online]. Available: https://www.littelfuse.com/~/media/electronics/datasheets/thermistor_probes_and_assemblies/littelfuse_thermistor_probes_assemblies_special_usp16673_datasheet.pdf.pdf. . [1st January 2020]

[41] (2019, Accessed: 30- Apr- 2020). ESP32-WROOM-32. 26. Available: https://www.espressif.com/sites/default/files/documentation/esp32-wroom-32_datasheet_en.pdf

[42] (2019). *UTF-8 string length & byte counter* [Online]. Available: https://mothereff.in/byte-counter. [Accessed: 20- Nov- 2019]

[43] *Interlink Electronics FSR Force Sensing Resistors*, 3rd ed ed. Digi-Key: interlink electronics, 2019.

[44] J. Perttunen, *Foot loading in normal and pathological walking*: University of Jyväskylä, 2002.

[45] (2020, Accessed 26 Jan. 2020). *FSR 402 Data Sheet (1st ed.)* [Online]. Available: https://www.trossenrobotics.com/productdocs/2010-10-26-DataSheet-FSR402-Layout2.pdf

[46] Ellisys, "Ellisys Bluetooth Video 6: BLE Power Consumption," ed: YouTube, 2018.